\documentclass[11pt,twoside]{article}
\usepackage{./asp2014}

\aspSuppressVolSlug
\resetcounters

\begin{document}

\title{Get ready for Gaia: cool white dwarfs in common proper motion with Tycho stars} 
\author{Nigel~Hambly, Nick Rowell and Marco Lam
\affil{Institute for Astronomy, SUPA (Scottish Universities Physics Alliance), School of Physics and Astronomy, University of Edinburgh, Royal Observatory, Blackford Hill, Edinburgh, UK; \email{nch@roe.ac.uk}}}

% This section is for ADS Processing.  There must be one line per author.
\paperauthor{Nigel~Hambly}{nch@roe.ac.uk}{}{Institute for Astronomy}{University of Edinburgh}{Edinburgh}{Scotland}{EH9~3HJ}{UK}
\paperauthor{Nick~Rowell}{nr@roe.ac.uk}{}{Institute for Astronomy}{University of Edinburgh}{Edinburgh}{Scotland}{EH9~3HJ}{UK}
\paperauthor{Marco~Lam}{mlam@roe.ac.uk}{}{Institute for Astronomy}{University of Edinburgh}{Edinburgh}{Scotland}{EH9~3HJ}{UK}

\begin{abstract}
We discuss the Gaia Data Release~1 (September 2016) 
and preliminary work on maximising the benefit for cool white dwarf (WD) science in
advance of the full parallax catalogue which will appear around one year
later in~DR2. The Tycho catalogue is used in conjunction with the 
all--sky ground based astrometric/photometric SuperCOSMOS Sky Survey in
order to identify candidate faint common proper motion objects to the
Tycho stars. Gaia~DR1 is supplemented by the Tycho--Gaia Astrometric Solution
catalogue containing some 2 million parallaxes with Hipparcos--like
precision for Tycho stars. While hotter, brighter WDs are present in 
Tycho, cooler examples are much rarer (if present at all) and CPM 
offers one method to infer precision distances for a statistically 
useful sample of these very faint WDs.
\end{abstract}

\section{Introduction}
ESA's Gaia mission has been in routine operations for two years. The first 
intermediate data release (GDR1) occurred in September 2016 and is based on the
first~14 months of data gathered by the spacecraft. Such a time baseline is
insufficient to disentangle reliably proper motion and parallax so the primary
data product of GDR1 will be a map of around one billion stars to V~$\sim20$
with mean epoch positions and single passband photometry but no proper motions
nor parallaxes. However, by using the Tycho data from the precursor Hipparcos
mission it is possible to anchor the positions for over~2 million brighter stars
(V~$<11.5$) at a mean early epoch around~1991 and hence solve for proper motion
and parallax. This yields trigonometric distances for $20\times$ as many stars as in the
Hipparcos catalogue with similar precision~\citep{2015A&A...574A.115M}. A bonus part of GDR1 
are these Tycho--Gaia Astrometric Solution (TGAS) sources. Data are released to
the community via a central Science Archive system and associated partner Data
Centers (e.g.~CDS) with extensive exploitation facilities and 
documentation\footnote{\url{http://archives.esac.esa.int/gaia/}}.

\section{White Dwarfs in Hipparcos and Tycho}

There are~10 or so WDs in the Hipparcos catalogue (which is at best complete to V~$\sim9$)
with parallaxes measured at a 5$\sigma$ level. The number of WDs present in the
Tycho catalogue (90\% complete to V~$\sim11.5$) is larger of course, so TGAS will
deliver in itself a major step forward for WD science via the addition of
precision distances (and therefore WD masses, radii, etc). However the TGAS WDs
will be biased towards the hot, bright end of the WD luminosity function simply
because of the bright magnitude limit of that catalogue coupled with the extreme
faintness of cooler WDs.

\section{White Dwarfs in common proper motion with Tycho stars}

One method for expanding significantly the sample of cool WDs with measured 
distances in GDR1 is to find those with high proper motion using existing
ground--based catalogues, e.g. the digitised optical all--sky Schmidt photographic
surveys, and associate them with Tycho stars via common proper motion. This can
be done with usefully high confidence given a sufficiently high lower proper
motion cut such that in any given part of the sky the likelihood of finding by
chance two nearby stars with the same high proper motion is negligibly small.
Figure~\ref{fig:rpm} illustrates this using the SuperCOSMOS Science 
Archive\footnote{\url{http://ssa.roe.ac.uk}} RECONS proper motion survey plus 
lower proper motion supplement~\citep{2004AJ....128..437H,2011MNRAS.417...93R} 
down to a lower proper motion
limit of~80~mas/yr. Wide binary candidates with separations of up to~16.7 arcmin
(1000~arcsec, corresponding to maximal separations of order 100,000~AU at 100~pc)
can be identified in this way and their position in reduced proper motion /
colour space shows some of them to be likely cool WDs. In Figure~\ref{fig:example} the image and
adjacent thumbnails show one example, an R~=~19.7 object with R--I~=~1.3 having
a proper motion within 1.5$\sigma$ of that of TYC~2734--750--1 (central in the main
picture) which is more than~8 magnitudes brighter in~R. Using common proper
motion it will be possible to infer accurately the distances to a large sample
of cool WDs prior to the availability of direct trigonometric parallaxes in
GDR2.

\articlefigure[scale=0.8,clip=true,viewport=0 0 418 500]{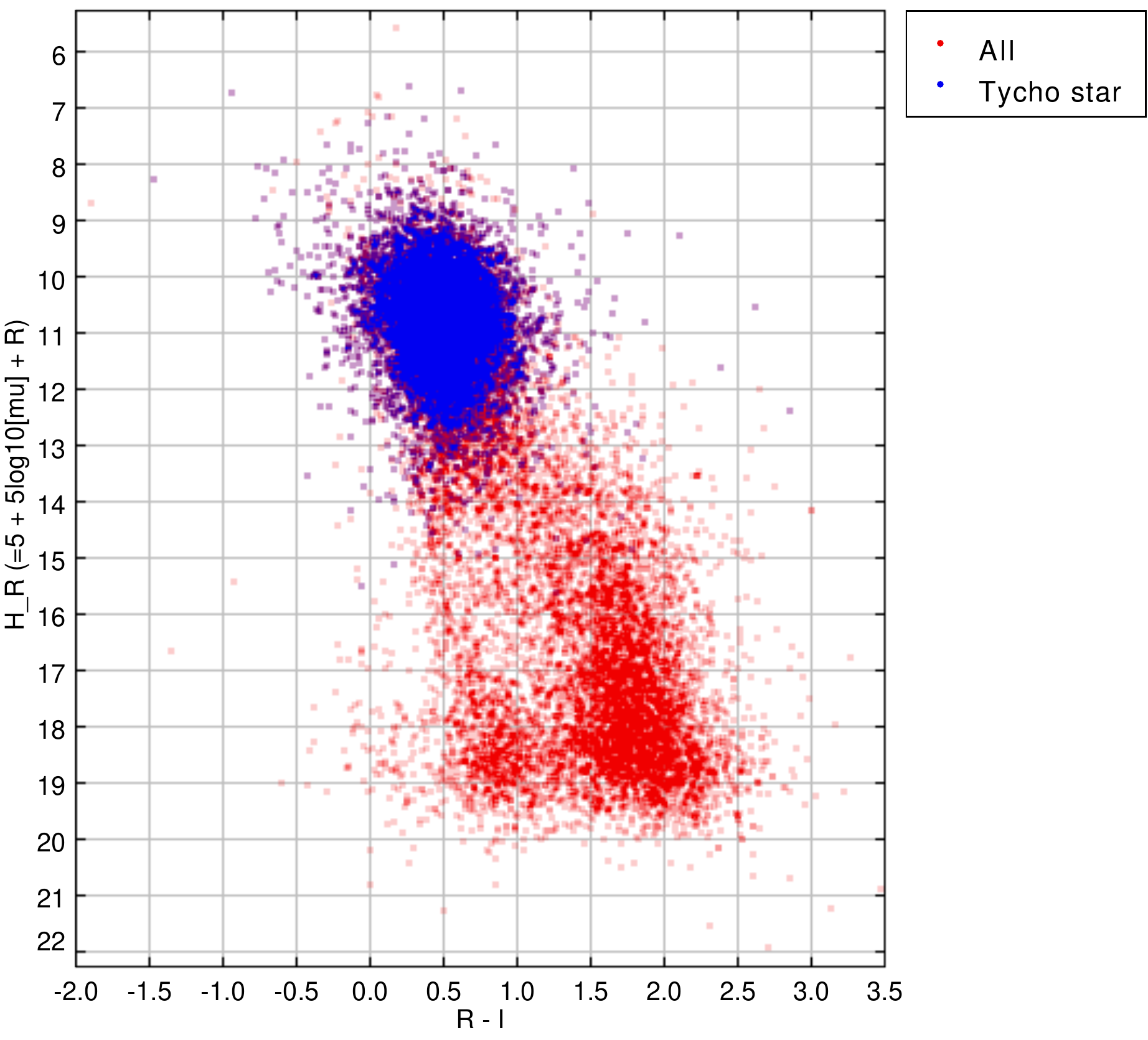}{fig:rpm}
{Reduced proper motion diagram for Tycho2 catalogue stars (blue) and 1.5$\sigma$ 
common proper motion objects from SuperCOSMOS (red). The usual dwarf, subdwarf 
and WD locii trace from the upper left to the lower right in this RPM--colour space. 
A search radius of~1000 arcsec was used along with a lower proper motion cut of~80~mas/yr. 
All the red objects share common proper motion with a Tycho star.}

\articlefigure[scale=0.4,angle=270]{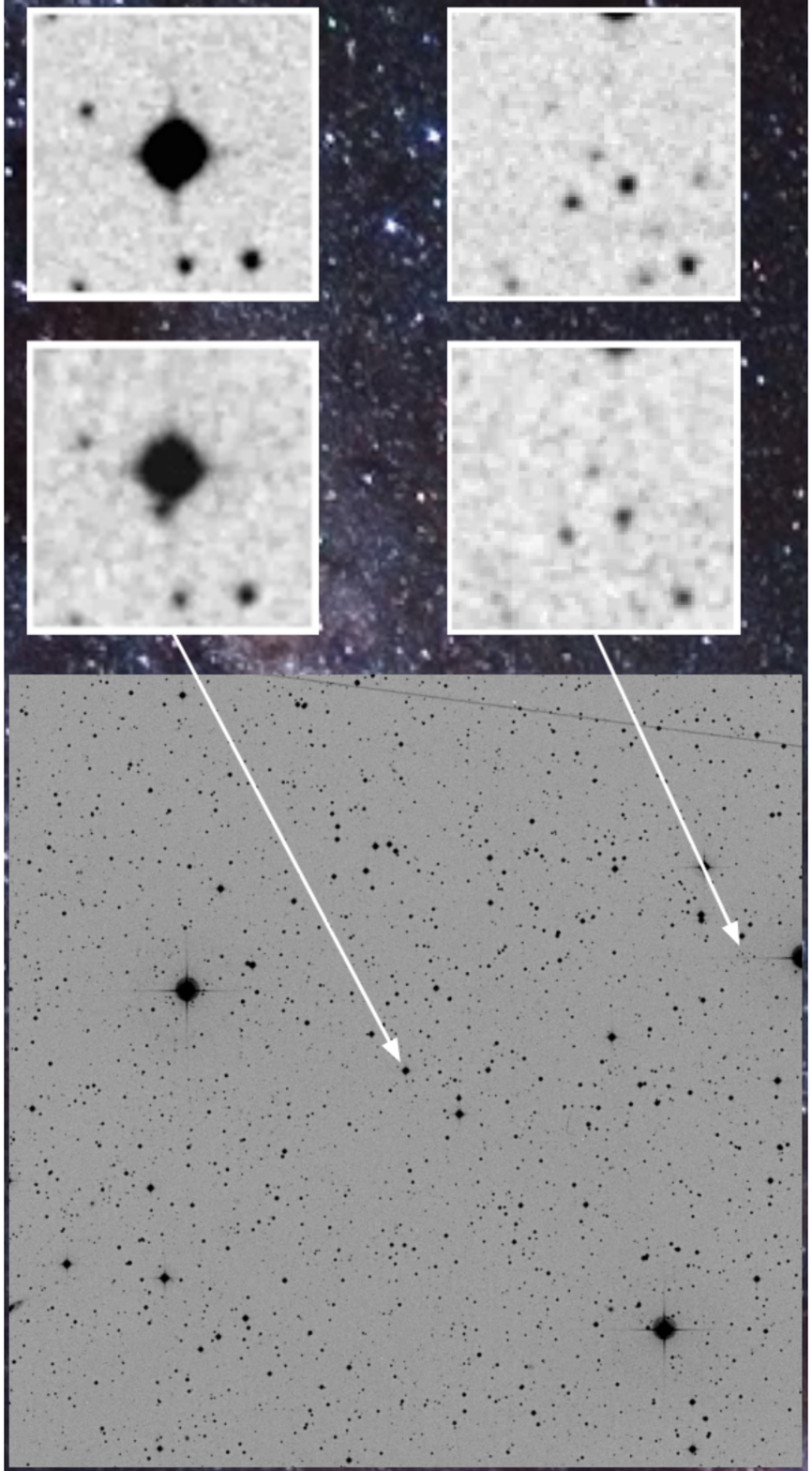}{fig:example}
{Example of a candidate wide binary identified via common  proper motion between
TYC~2734--750--1 (R~=~11.3) and a cool WD (R~=~19.7; R--I~=~1.3). The wide angle
image is~30 arcmin on each side while the thumbnails are~1 arcmin. Thumbnails on
the left are from the POSS--E plate scan while those on the right are from the
2nd epoch red survey plate (exposed some~40 years later).}

\section{White dwarfs in GDR2 and subsequent releases}

Of course the major leap forward for the science of all WD fields of study will
be from GDR2 onwards with the availability of full five--parameter
astrometric solutions and (some) radial velocities~\citep{2007ASPC..372..139J}. Even then full 6D kinematics
will not be available for the majority of isolated cool WDs because of the
unknown and generally unmeasurable radial velocity component. Moreover accurate
model--independent ages are difficult to establish for such objects which leads
to, for example, poor coverage in the Initial--Final Mass Relation. If, however, 
a cool WD can be
associated with a normal star via common proper motion then the distance and
radial velocity will nearly always be more accurately measured (or indeed may
only be measurable) for the system using the brighter component. If the system
can be identified kinematically with a cluster, association or moving group of
known age and metallicity etc.\ then so much the better and hence the technique
of `benchmarking' systems~\citep{2013EPJWC..4706002G,2012ApJ...746..144Z} identified via common proper motion will
remain a valuable tool for WD science in the Gaia era.

An interesting aspect of this work is in the implementation details of the
CPM search algorithm. Catalogue pairing is, of course, a solved problem in 
computer science via `plane sweep' algorithms that deliver true $O(N\log M)$
performance~(see~\citealt{2005ASPC..347..346D} and references therein) for
small search radii. For the larger search radii needed in this application,
we have further refined the implementation using the `zoned join' technique
developed by~\cite{gray} where the input catalogue(s) are first split into
separate Declination zones of extent equal to the maximum search radius 
required to identify CPM wide binaries (in this case 1000~arcsec). This is
particularly important when scaling up to a self--join of a billion--row
scale catalogue, as will be the case when GDR2 is released.

\acknowledgements This research has made use of data obtained from the SuperCOSMOS Science Archive, prepared and hosted by the Wide Field Astronomy Unit, Institute for Astronomy, University of Edinburgh, which is funded by the UK Science and Technology Facilities Council.

%\bibliography{editor}  % For BibTex

% For non-BibTex:

\end{document}